\documentclass[10pt,a4paper]{amsart}
\usepackage[latin1]{inputenc}
\usepackage{mathrsfs}
\usepackage{dsfont}
\usepackage{amsmath}
\usepackage{amssymb}
\usepackage{amsthm}
\usepackage{amsfonts}
\usepackage{amstext}
\usepackage{amsopn}
\usepackage{amsxtra}
\usepackage{mathrsfs}
\usepackage{dsfont}
\usepackage{esint}
\usepackage{pst-all}
\usepackage{graphicx}
\usepackage{bbding}
\newtheorem{theorem}{Theorem}

\newtheorem{lemma}{Lemma}

\newtheorem{definition}{Definition}

\newcommand{\ii}{\infty}
\newcommand\R{{\ensuremath {\mathbb R} }}
\newcommand\C{{\ensuremath {\mathbb C} }}

\newcommand\1{{\ensuremath {\mathds 1} }}
\renewcommand\phi{\varphi}
\newcommand{\gH}{\mathfrak{H}}

\newcommand{\wto}{\rightharpoonup}

\newcommand{\cD}{\mathcal{D}}

\newcommand{\br}{\mathbf{r}}
\newcommand{\bp}{\mathbf{p}}

\newcommand{\alp}{\boldsymbol{\alpha}}
\newcommand{\dnr}{{\rm d}^Nr}
\newcommand{\dr}{{\rm d}r}
\newcommand{\dddr}{{\rm d}^3r}
\renewcommand{\epsilon}{\varepsilon}
\newcommand\pscal[1]{{\ensuremath{\left\langle #1 \right\rangle}}}

\newcommand{\Spec}{{\rm Spec}\;}

\renewcommand{\geq}{\geqslant}
\renewcommand{\leq}{\leqslant}

\numberwithin{equation}{section}

\begin{document}
%%%%%%%%%%%%%%%%%%%%%%%%%%%%%%%%%%%%%%%%%%%%%
%%%%%%%%%%%%%%%%%%%%%%%%%%%%%%%%%%%%%%%%%%%%%

\title[Spurious Modes in Dirac Calculations]{Spurious Modes in Dirac Calculations and How to Avoid Them}

\author[M. Lewin]{Mathieu LEWIN}
 \address{CNRS \& Laboratoire de Mathématiques (CNRS UMR 8088), Universit\'e de Cergy-Pontoise, F-95000 Cergy-Pontoise Cedex, France.}
  \email{mathieu.lewin@math.cnrs.fr}

\author[\'E. S\'er\'e]{\'Eric S\'ER\'E}
\address{Ceremade (CNRS UMR 7534), Universit\'e Paris-Dauphine, Place de Lattre de Tassigny, F-75775 Paris Cedex 16, France.}
\email{sere@ceremade.dauphine.fr}

\date{\today. \scriptsize~\copyright~2013 by the authors. This paper may be reproduced, in its entirety, for non-commercial~purposes. To be published in the book \emph{Many-Electron Approaches in Physics, Chemistry and Mathematics: A Multidisciplinary View} edited by Volker Bach and Luigi Delle Site.}

\begin{abstract}
In this paper we consider the problem of the occurrence of spurious modes when computing the eigenvalues of Dirac operators, with the motivation to describe relativistic electrons in an atom or a molecule. We present recent mathematical results which we illustrate by simple numerical experiments. We also discuss open problems.
\end{abstract}

\maketitle

%\tableofcontents

Computing the eigenvalues of an operator on a computer can be a subtle task, in particular when one is interested in those lying in a gap of the spectrum. In this case, \emph{spurious modes} can sometimes appear and persist when the size of the discretization basis is increased. The phenomenon, called \emph{spectral pollution}, is well-known and well documented. For instance, it is encountered when dealing with perturbations of periodic Schrödinger operators~\cite{BouLev-07,CanEhrMad-12} or Sturm-Liouville operators~\cite{StoWei-93,StoWei-95,AceGheMar-06}. It also appears in elasticity, electromagnetism and hydrodynamics~\cite{CseSil-70,Bossavit-90,SchWol-94,JiaWuPov-96,RapSanSanVas-97,BunDav-99,BofBreGas-00,FerRaf-02}.

In this paper we are interested in relativistic computations based on the Dirac operator, like those used in quantum chemistry and atomic physics. The spectrum of the free Dirac operator is $(-\ii,-mc^2]\cup[mc^2,\ii)$ and adding an external potential usually creates eigenvalues in the gap $(-mc^2,mc^2)$. 
Computing them might lead to spurious modes. Practical solutions to overcome this problem have been proposed a long time ago~\cite{DraGol-81,Grant-82,Kutzelnigg-84,StaHav-84,DyaFae-90,Pestka-03,Shaetal-04,BouBou-10}, the most famous of them being the \emph{kinetic balance method}. Until recently, these methods had not been studied from a mathematical perspective. The purpose of this paper is to review and illustrate the results of our article~\cite{LewSer-10}, where we rigorously investigated the validity of these methods. In particular, we show under which precise condition the kinetic balance prescription is guaranteed to avoid spurious eigenvalues. Several open problems remain, however, and we will discuss them as well.

Relativistic effects were almost always neglected in quantum chemistry calculations, until it was realized in the 1970s that they are actually very important to account for some elementary properties of heavy atoms. The problem of spurious modes can in principle appear in any calculation based on the Dirac operator. For a general presentation of the Dirac equation from the point of view of quantum chemistry, we refer to~\cite{Schwerdtfeger1} and to the chapter of B. Simmen and M. Reiher in this book. We remark that, in applications of Density Functional Theory, relativistic effects are rarely considered. They are often implicitly included into pseudo-potentials of the nuclei which includes the inner (relativistic) electrons (see, in particular, the chapters of A.T.~Tzanov and M.E.~Tuckerman, of L.M.~Ghiringhelli, and of O.A.~von Lilienfeld).

%%%%%%%%%%%%%%%%%%%%%%%%%%%%
\section{What is Spectral Pollution?}
%%%%%%%%%%%%%%%%%%%%%%%%%%%%

In this section we quickly review some general properties of spectral pollution, with an emphasis on the Dirac case. 

%%%%%%%%%%%%%%%%%
\subsection{Self-adjointness, Domains and all that}

In quantum mechanics, we have to manipulate \emph{self-adjoint operators} $A$, which have a real spectrum and for which Schrödinger's equation $i\hbar\partial_t\psi=A\psi(t)$ has a unique solution, by Stone's theorem. In infinite dimension, the concept of a self-adjoint operator is not always easy~\cite{Simon-85,Sutcliffe-08}. Finding a self-adjoint realization of an operator $A$ in a Hilbert space\footnote{In our examples we will have $\gH=L^2(\Omega)$, the space of square-integrable functions on a domain $\Omega$ in the $N$-dimensional space $\R^N$. We will encounter two main cases: that of the whole physical space $\Omega=\R^3$ and that of the half line $\Omega=(0,\ii)$ useful to deal with radial functions.} $\gH$ amounts to choosing a \emph{domain} $\cD(A)\subset\gH$ on which $A$ is well-defined and has certain good properties that we do not give in detail here~\cite{Davies-98}.

In the good situations (namely when $A$ is \emph{essentially self-adjoint} on a natural subspace) there is no ambiguity for $\cD(A)$ and this is the case for most perturbations of differential operators in $\R^N$. When $\gH=L^2(\Omega)$, with $\Omega$ an open bounded subset in $\R^N$, then $\cD(A)$ should include boundary conditions and a choice has to be made. This is of course important as the spectrum of $A$, which is our primary interest here, depends on these boundary conditions.

Let us now give two examples. In the non-relativistic case we have $A=-\hbar^2\Delta/(2m)$ where
$\Delta$ is the Laplace operator and $\gH=L^2(\R^3)$, the space of square-integrable functions on $\R^3$. We then take 
$$\cD\big(-\hbar^2\Delta/(2m)\big)=\left\{\psi:\R^3\to\C\ \Big\vert\ \int_{\R^3}\big(|\psi(\br)|^2+|\Delta \psi(\br)|^2\big)\,\dddr\ \text{is finite}\right\}$$
which is a \emph{Sobolev space} often denoted as $H^2(\R^3)$. The assumption that $\Delta \psi$ is square-integrable is mandatory to ensure that $A$ maps functions in the domain $\cD(A)$ into the ambient Hilbert space $\gH=L^2(\Omega)$. The spectrum of the Laplacian on this domain is the half line
$$\Spec\big(-\hbar^2\Delta/(2m)\big)=[0,\ii).$$
There is no eigenvalue in this spectrum. Namely there does not exist any square-integrable function $\psi$ such that $-\hbar^2\Delta/(2m) \psi=\lambda \psi$. There only exist approximates eigenvectors, which means a sequence $(\psi_n)_{n\geq1}$ such that $\int_{\R^N}|\psi_n|^2=1$ and $-\hbar^2\Delta/(2m) \psi_n-\lambda \psi_n\to0$ as $n\to\ii$.\footnote{Take for instance $\psi_n(\br)=\exp(i\bp\cdot\br/\hbar)n^{-N/2}\chi(\br/n)$ for some smooth $\chi$ with $\int_{\R^N}|\chi(\br)|^2\dnr=1$ and a momentum $\bp$ such that $p^2=2m\lambda$.} 
In this special situation, one speaks of \emph{continuous spectrum}.

\begin{figure}[h]
\small
\input{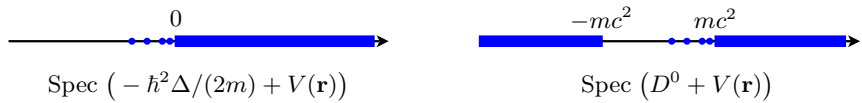}
\caption{The typical spectrum of the non-relativistic Schrödinger operator (left) and the Dirac operator (right), in an external potential $V(\br)$.\label{fig:comparison}}
\end{figure}

If we add an electric potential $V(\br)$ to our kinetic energy operator $-\hbar^2\Delta/(2m)$, and if $V(\br)$ is smooth enough and decays at infinity, then the domain of $-\hbar^2\Delta/(2m)+V(\br)$ will be the same as for $V\equiv0$. The spectrum will still contain the half line $[0,\ii)$. Negative eigenvalues can appear if $V$ is sufficiently negative in some part of space, corresponding to bound states of the system. They all have a finite multiplicity, and they can only accumulate at 0 (Figure~\ref{fig:comparison}).

For relativistic particles, one has to use the Dirac operator, which acts on $4$-spinors, that is, on square-integrable functions on $\R^3$ taking values in $\C^4$. It is given by
$$D^0=-ic\hbar \sum_{k=1}^3\alpha_k\, \frac{\partial}{\partial x_k}+mc^2\beta,$$
and its domain of definition is now the Sobolev space  
$$\cD(D^0)=H^1(\R^3):=\left\{\Psi:\R^3\to\C^4\ \Big|\ \int_{\R^3}\big(|\Psi(\br)|^2+\left|\nabla \Psi(\br)\right|^2\big)\,\dddr\ \text{is finite}\right\}.$$
Its spectrum is the union of two intervals,
$$\Spec(D^0)=(-\ii,-mc^2]\cup[mc^2,\ii),$$
which follows from charge-conjugation symmetry. It is again a purely continuous spectrum, without any eigenvalue. If we add an external electric potential $V(\br)$ which is smooth and decays at infinity, then the domain $\cD(D^0+V)$ and the essential spectrum do not change. Eigenvalues can appear in the gap $(-mc^2,mc^2)$ (see Figure~\ref{fig:comparison}), and we are interested in computing them numerically.

The situation is more subtle when the potential is the one generated by a pointwise nucleus (say of charge $eZ$):
$$V(\br)=-\frac{e^2Z}{r},$$
see~\cite{Thaller}.
The domain of $D^0+V$ is again the same as for $D^0$, provided $e^2Z\leq \hbar c \sqrt{3}/2$. The spectrum then contains a sequence of positive eigenvalues in the gap, converging to $mc^2$. For $\hbar c \sqrt{3}/2\leq e^2Z\leq \hbar c$ the domain is different and contains a further boundary condition at the origin. For $e^2Z> \hbar c$, there are infinitely many possibilities for $\cD(D^0+V)$ none of which  seems to have a particular physical meaning. In order to simplify our exposition, we always assume for simplicity that $e^2Z\leq \hbar c \sqrt{3}/2$, so that $\cD(D^0+V)=\cD(D^0)=H^1(\R^3)$.
We also choose a system of units such that $m=c=\hbar=1$. We are therefore only left with $\alpha=e^2$, the coupling constant which must satisfy $\alpha Z\leq \sqrt{3}/2$.

In a central potential we can look at the restriction of $D^0+V$ to a particular symmetry subspace. For example, in the sector of total angular momentum $j=1/2$, azimuthal angular momentum $j_z=0$ and spin orbit number $\kappa=-1$ (in which lies the ground state), the wave functions take the special form
$$\Psi(\br)=\frac{u(r)}r\begin{pmatrix}1\\0\\0\\0\end{pmatrix}+\frac{v(r)}{r}\begin{pmatrix}0\\0\\ \frac{1}{\sqrt3}Y^0_1(\omega)\\-\frac{\sqrt2}{\sqrt3} Y^1_1(\omega)\end{pmatrix},$$
where $\omega=\br /r$ is the angular part of $r$ and the Dirac eigenvalue equation becomes
\begin{equation}
\left(
\begin{array}{c|c}
\displaystyle1-\frac{\alpha Z}{r} & \displaystyle-\frac{\rm d}{{\rm d} r}-\frac{1}{r}\\[0.2cm]
\hline
\displaystyle\frac{\rm d}{{\rm d} r}-\frac{1}{r} & \displaystyle-1-\frac{\alpha Z}{r}
\end{array}\right)
\begin{pmatrix}u\\ v\end{pmatrix}
= \lambda \begin{pmatrix}u\\ v\end{pmatrix}
\label{eq:radial_basis}
\end{equation}
in the Hilbert space $L^2(\R^+,\dr)$. Expressed in terms of the functions $u$ and $v$, the domain becomes
\begin{equation*}
\bigg\{u,v:\R^+\to\C\ \Big|\ \int_0^\ii\Big(|u(r)|^2+|v(r)|^2+|u'(r)|^2+|v'(r)|^2\Big)\,\dr \text{ is finite}\bigg\}. 
\end{equation*}

\subsection{Approximating the Spectrum}

To find an approximation on a computer of the eigenvalues of the Dirac operator in an electrostatic potential $V(\br)$,
$$D^V:=D^0+V(\br),$$ 
we choose a finite-dimensional space $W\subset\cD(D^0+V)=H^1(\R^3)$, and we compute the matrix of the restriction of $D^V$ to $W$. Simply, if $b_1(\br),...,b_d(\br)$ is a basis of $W$, then the associated $d\times d$ matrix is $(D^V)_{|W}=(\pscal{b_i,D^Vb_j})_{1\leq i,j\leq d}$, where $d$ is the dimension of $W$. Its eigenvalues now solve the generalized eigenvalue equation
\begin{equation}
(D^V)_{|W}x=\lambda Sx,
\label{eq:discretized} 
\end{equation}
where $S=(\pscal{b_i,b_j})_{1\leq i,j\leq d}$ is the overlap matrix. Here and elsewhere we use the notation 
$$\pscal{\Psi,\Phi}=\int_{\R^3}\Psi(\br)^* \Phi(\br)\,\dddr=\sum_{j=1}^4\int_{\R^3}\overline{\Psi(\br)_j} \Phi(\br)_j\,\dddr$$ 
to denote the ambient scalar product for $4$-spinors. We have assumed that $W\subset\cD(D^V)=H^1(\R^3)$ which guarantees that $\pscal{b_i,D^Vb_j}$ makes sense, but this is not the optimal condition. The scalar product $\pscal{b_i,D^Vb_j}$ is usually well-defined on a larger space called the \emph{quadratic form domain} of $D^V$, but we do not discuss this further, for simplicity. 

Having found the spectrum of the $d\times d$ matrix $(D^V)_{|W}$, we want to know if the obtained eigenvalues are good approximations to the elements of the spectrum of $D^V$. This approximation must improve when the size of the basis grows and, for this reason, it is customary to instead consider a sequence of discretization spaces $W_n$, such that  $\dim W_n\to\ii$, and ask whether the approximate eigenvalues converge to the true ones as $n\to\ii$.

It is clear that if we hope for a good representation of the eigenfunctions of $D^V$, then the approximation sequence $W_n$ must be adapted to $D^V$ in some way. One condition is that $W_n$ approximates the domain $H^1(\R^3)$ of $D^V$. This means that for any $\Psi\in H^1(\R^3)$, there exists an approximating sequence $(\Psi_n)_{n\geq1}\subset H^1(\R^3)$ with $\Psi_n\in W_n$ such that 
\begin{equation}
\lim_{n\to\ii}\int_{\R^3}\left(|\Psi_n(\br)-\Psi(\br)|^2+|\nabla(\Psi_n-\Psi)(\br)|^2\right)\dddr=0.
\label{eq:dense} 
\end{equation}
This completeness condition is satisfied for most approximation schemes, like the finite element method for instance. In the paper~\cite{KlaBin-77b}, Klahn and Bingel provided some simple conditions (based on the so-called Müntz theorem) which imply that~\eqref{eq:dense} is satisfied for a basis made of gaussian functions, as is used in most quantum chemistry programs.

It is well-known that the condition~\eqref{eq:dense} ensures that we ind the whole spectrum of $D^V$ in the limit of a large basis set (see, e.g., \cite[Prop. 2]{BouBouLew-12}):

\begin{theorem}[The spectrum is well-approximated]
If $W_n$ approximates the Sobolev space $H^1(\R^3)$ in the sense of~\eqref{eq:dense}, then, for any $\lambda$ in the spectrum of $D^V$, there exists $\lambda_n$ in the spectrum of $(D^V)_{|W_n}$ converging to $\lambda$ as $n\to\ii$. Similarly, any non-degenerate eigenfunction of $D^V$ is approximated in $H^1(\R^3)$ by an eigenfunction of $(D^V)_{|W_n}$ in the limit $n\to\ii$.
\end{theorem}

Since $(D^V)_{|W_n}$ is a finite matrix, an eigenfunction is here just an eigenvector of this matrix. Another equivalent definition is given in~\eqref{eq:approx_eigenfn} below.

\subsection{Spurious eigenvalues}

That we are sure to get the spectrum of $D^V$ in the limit of a large basis set does not mean at all that we are in a good situation. Indeed, it can happen that in the limit we get much more than only the spectrum of $D^V$, and this is precisely what spectral pollution is about. We can give a precise definition of a spurious eigenvalue as follows:

\begin{definition}[Spurious spectrum]
A real number $\lambda\in(-1,1)$ is called a spurious eigenvalue of $D^V$ (relative to the approximation scheme $W_n$), if there exists $\lambda_n$ in the spectrum of $(D^V)_{|W_n}$ converging to $\lambda$ as $n\to\ii$, such that 

\smallskip

\noindent$\bullet$ \textit{either} $\lambda$ is not in the spectrum of $D^V$;

\smallskip

\noindent$\bullet$ \textit{or} $\lambda$ is an isolated eigenvalue of finite multiplicity $M$ of $D^V$, but its multiplicity is overestimated in the limit $n\to\ii$. This means that there are more than $M$ eigenvalues of $(D^V)_{|W_n}$ counted with multiplicity in the interval $(\lambda-\epsilon_n,\lambda+\epsilon_n)$, for some $\epsilon_n\to0$.
\end{definition}

In practice one calls $\lambda_n$ the spurious mode instead of its limit $\lambda$ (but in principle the limit should be taken to be sure that the spurious mode persists).

In order to clarify the situation, we will now immediately give two simple examples of spurious eigenvalues. We start with an academic example, before turning to the Dirac operator in a Coulomb potential.

\subsubsection*{An academic example}
We take $\gH=L^2(0,2\pi)$ as Hilbert space and recall the Fourier basis $\{1,\cos(nr),\sin(nr)\}_{n\geq1}$. Any function in $\gH$ can be expanded in this basis as follows,
$$f(r)=\frac{a_0}{\sqrt{2\pi}}+\frac{1}{\sqrt{\pi}}\sum_{n\geq1}a_n\,\cos(nr)+b_n\,\sin(nr),$$
where
$$\int_{0}^{2\pi} |f(r)|^2\,\dr=|a_0|^2+\sum_{n\geq1}|a_n|^2+|b_n|^2.$$
We now introduce the orthogonal projection $P$ onto the odd modes,
$$(Pf)(r)=\frac{1}{\sqrt{\pi}}\sum_{n\geq1}b_n\,\sin(nr).$$
The operator $P$ is bounded and hence can be defined on the whole space $\cD(P)=L^2(0,2\pi)$, there is no subtlety of domain for $P$. The operator $P$ is diagonal in the Fourier basis, which are thus its eigenvectors. Its spectrum is simply 
$$\Spec(P)=\{0,1\}$$ 
where the two eigenvalues $0$ and $1$ have an infinite multiplicity. 

Now we choose our approximation space $W_n$ by picking all the even and odd modes less or equal than $n-1$, and mixing the two $n$ modes as follows:
\begin{multline*}
 W_n={\rm span}\big\{1,\sin(r),\cos(r),...\\
...,\sin((n-1)r),\cos((n-1)r),\cos(\theta)\cos(nr)+\sin(\theta)\sin(nr)\big\}.
\end{multline*}
This is of course very artificial but it helps to understand the phenomenon of spectral pollution in more practical situations.
The matrix of $P_{|W_n}$ in this basis is
$$P_{|W_n}=\left(\begin{matrix}
0 & & & & &\\
& 1 &  & & &\\
 & & 0 & & &\\
 & & & 1 & &\\
& & & & \ddots &\\
 & & & & & \sin^2(\theta)
\end{matrix}\right)$$
and thus 
$$\Spec (P_{|W_n})=\{0,\sin^2(\theta),1\}$$
for all $n$. The eigenvalue $\sin^2(\theta)$ persists in the limit $n\to\ii$ and it is spurious. The corresponding eigenfunction is $\pi^{-1/2}\sin(nr)$ which oscillates very fast. Of course, by mixing several modes in the same way, we can create an arbitrary number of spurious modes, having any value in the gap $(0,1)$. By taking a number of spurious modes tending to infinity, we can even fill the whole interval $(0,1)$ with spurious eigenvalues.

This academic example reveals most of the nature of spectral pollution. A spurious mode is obtained when states from the spectrum above and below the considered gap are mixed together. It is because there are infinitely many states above and below that this can happen for a large basis set. The corresponding spurious eigenfunction will usually behave badly. It will oscillate very fast, or concentrate at the boundary of the domain, for instance.

Before turning to an example involving the Dirac operator, let us make an important remark. As we have explained, spurious modes appear in gaps of the essential spectrum, because of the two infinite-dimensional ``reservoirs'' below and above the gap. Spurious modes will \emph{never} appear below or above the essential spectrum, when the considered operator is bounded from below or from above. This claim can be proved by using the well-known min-max characterization of eigenvalues, which is usually referred to as the Hylleraas-Undheim-MacDonald (HUM) theorem in the quantum chemistry literature~\cite{HylUnd-30,McDonald-33}, and as the Rayleigh-Ritz variational principle in mathematics. This principle does not apply to eigenvalues in gaps. There exists a min-max characterization of the eigenvalues in gaps~\cite{DolEstSer-00,EstLewSer-08} but it is much more complicated and it does not prevent the occurrence of spurious modes in general.

\subsubsection*{A Numerical Example with the Dirac Operator}
We can now provide a more practical example involving the (radial) Dirac operator. We restrict ourselves to the sector of total angular momentum $j=1/2$ and spin-orbit $\kappa=-1$ mentioned before in~\eqref{eq:radial_basis}, and we choose a basis made of gaussians, for the radial parts $u(r)$ and $v(r)$. We take the same basis for $u(r)$ and $v(r)$, we do not impose any kinetic balance as we will later do in Section~\ref{sec:KB}. To this basis, we add a vector which is a mixture of an upper and lower spinor, in the same spirit as in the previous example:
\begin{multline}
W_n=\left\{e^{-a_1r^2}\begin{pmatrix}1\\0\\0\\0\end{pmatrix}\,,\,e^{-a_1r^2}\begin{pmatrix}0\\0\\ \frac{1}{\sqrt3}Y^0_1(\omega)\\-\frac{\sqrt2}{\sqrt3} Y^1_1(\omega)\end{pmatrix}\, ,...,\,e^{-a_nr^2}\begin{pmatrix}1\\0\\0\\0\end{pmatrix}\,,\right.\\
\left. e^{-a_nr^2}\begin{pmatrix}0\\0\\ \frac{1}{\sqrt3}Y^0_1(\omega)\\-\frac{\sqrt2}{\sqrt3} Y^1_1(\omega)\end{pmatrix}\;,\; \cos\theta e^{-br^2}\begin{pmatrix}1\\0\\0\\0\end{pmatrix}\,+\,\sin\theta e^{-br^2}\begin{pmatrix}0\\0\\ \frac1{\sqrt3}Y^0_1(\omega)\\-\frac{\sqrt2}{\sqrt3} Y^1_1(\omega)\end{pmatrix}\right\}
\label{eq:6-31G-no-kinetic}
\end{multline}
where $a_1,...,a_n$ are the coefficients of the (uncontracted) gaussians of the 6-31G basis for Zinc ($Z=30$) given in Table~\ref{tab:6-31G}.

\begin{table}[h]
\begin{tabular}{lllll}
82400.940 & 12372.550 & 2818.3510 & 1732.5690 & 794.57170 \\
412.71490 & 254.72320 & 133.67800 & 87.138800 & 69.364920 \\
50.385850 & 23.620820 & 20.583580 & 10.184710 & 8.5059400 \\ 
4.3340820 & 2.8238420 & 1.8109180 & 1.0395430 & 0.7148410 \\  
0.1432640 & 0.0492960
\end{tabular}
\caption{The coefficients $a_1\alpha^{-2}<\cdots<a_n\alpha^{-2}$ of the 6-31G basis set for $Z=30$ and $n=22$.\label{tab:6-31G}}
\end{table}

\begin{figure}[h]
\includegraphics[width=10cm]{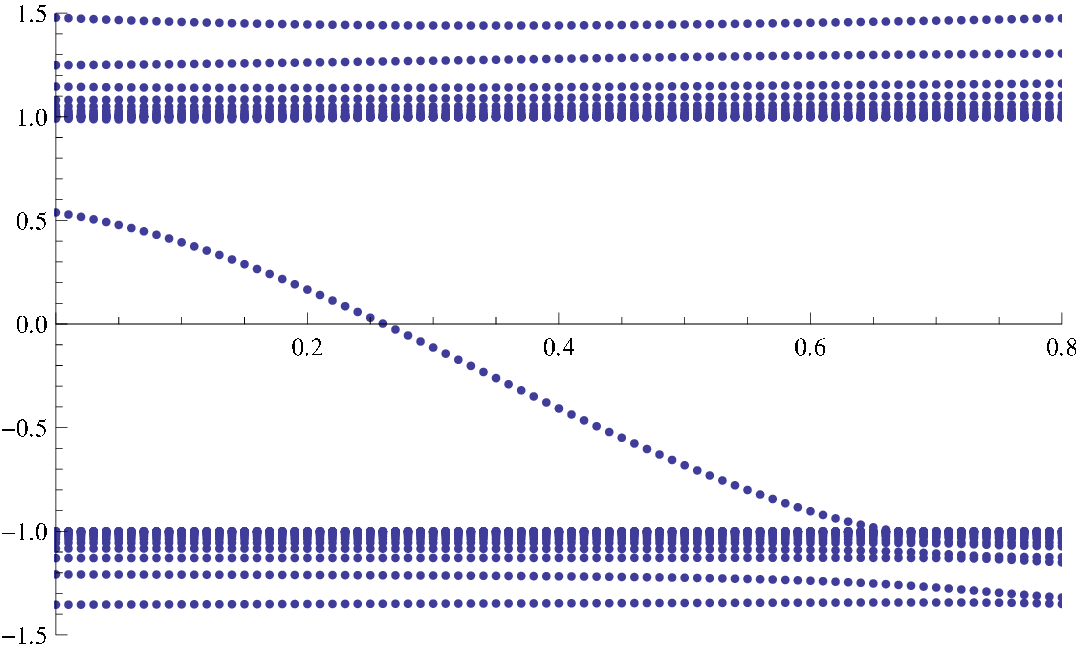}
\caption{Spectrum of $D^0-30\alpha/r$ computed in the basis set~\eqref{eq:6-31G-no-kinetic} and plotted vertically in terms of the parameter $\theta$.\label{fig:no-kinetic}}
\end{figure}

In Figure~\ref{fig:no-kinetic} we show the spectrum of the Dirac operator $D^0-30\alpha/r$ computed in the basis set~\eqref{eq:6-31G-no-kinetic}, with $b=10^6\alpha^2$ and as a function of the mixing parameter $\theta$. We notice the presence of a spurious mode which varies a lot when $\theta$ is changed. The true ground state energy is 
$$\lambda_1^{\rm true}=\sqrt{1-(30\alpha)^2}\simeq0.975729$$ 
and, without the additional mode, its 6-31G approximation is found to be $\lambda_1^\text{app}\simeq0.975739$. With the additional spurious mode, the value of the approximate ground state energy deteriorates to $\lambda_1^{\rm spu}\simeq 0.996578$ at $\theta=0.5$. This decrease of quality in the approximation for the ground state eigenvalue is a clear motivation to construct a better basis set.

%%%%%%%%%%%%%%%%%
\subsection{Weak limit of spurious eigenvectors}

We have seen that there can be spurious eigenvalues in Dirac calculations, and we have given a simple example of such a phenomenon. Here we quickly discuss an important property of spurious eigenvectors.

Consider a sequence of approximation spaces $W_n$ and assume that $\lambda\notin\Spec(D^V)$ is a spurious eigenvalue. Then there is a solution to the eigenvalue equation $(D^V)_{|W_n}x_n=\lambda_nS_nx_n$ in $W_n$, for some sequence of spurious eigenvalues $\lambda_n\to\lambda$. Introducing the corresponding \emph{approximate eigenfunction} $\Psi_n(\br)=\sum_{j=1}^{d_n} (x_n)_j\,b_j(\br)$ in $W_n$ with $\int_{\R^3}|\Psi_n(\br)|^2\dddr=1$, this means that we have
\begin{equation}
 \int_{\R^3}\Phi_n(\br)^\ast\, \big(D^0+V(\br)-\lambda_n\big)\Psi_n(\br)\,\dddr=0,\quad\text{for all $\Phi_n\in W_n$.}
\label{eq:approx_eigenfn}
\end{equation}
We recall that $\Psi_n$ is said to \emph{weakly converge to 0} if $\int_{\R^3}\Phi(\br)^* \Psi_n(\br)\dddr\to0$,
for any fixed $\Phi\in L^2(\R^3)$. In other words, it becomes asymptotically orthogonal to any fixed state $\Phi$ in the limit $n\to\ii$.

The following is an important property of spurious eigenvectors.

\begin{lemma}[Spurious eigenvectors weakly tend to 0]
If $\lambda\notin\Spec(D^V)$ is a spurious eigenvalue as above, then we must have $\Psi_n\wto0$ weakly in $L^2(\R^3)$.
\end{lemma}

The proof of the lemma is elementary. First, we use that $D^V$ is symmetric:
$$\int_{\R^3}\Phi_n(\br)^\ast\, \big(D^0+V(\br)-\lambda_n\big)\Psi_n(\br)\,\dddr=\pscal{(D^V-\lambda)\Phi_n,\Psi_n}=0.$$
By the approximation property~\eqref{eq:dense} of $W_n$ we know that we can approximate any function $\Phi\in H^1(\R^3)$, that is we can find a sequence $\Phi_n\in W_n$ such that $D^V\Phi_n\to D^V \Phi$. On the other hand, since $\int|\Psi_n|^2=1$ for all $n$, we know that $\Psi_n$ admits a subsequence which  weakly converges to some $\Psi$. Passing to the limit we get $\pscal{(D^V-\lambda)\Phi,\Psi}=0$. But this is true for all $\Phi\in H^1(\R^3)$ and this now implies $(D^V-\lambda) \Psi=0$. Since $\lambda$ is not in the spectrum of $D^V$ by assumption, then we must have $\Psi\equiv0$. We have proved that the limit of any weakly convergent subsequence is zero. This says that $\Psi_n\wto0$ weakly, and the proof is finished.

The result requires to have $\lambda\notin\Spec(D^V)$. As we said there is another type of spurious modes corresponding to a $\lambda$ which belongs to the true spectrum, but whose multiplicity is over-estimated. This situation is more complicated~\cite{BouBouLew-12} and we do not consider it here. Indeed, this almost never happens in practice. As can be seen from the numerical experiments, spurious modes are usually very unstable: they tend to move a lot when the parameters of the basis are changed, contrary to the other eigenvalues of the discretized spectrum. Typically, spurious modes will therefore not end up exactly on a true eigenvalue of $D^V$.

%%%%%%%%%%%%%%%%%
\subsection{How to identify the spurious spectrum?}\label{sec:method}

In this section we discuss a simple strategy to construct spurious modes, which does not rely on any chosen approximate basis set. The method is based on the previous remark that spurious eigenvectors necessarily tend to zero weakly. 

Suppose that we can construct a sequence $\Psi_n$ of normalized functions, such that 
\begin{enumerate}
\item $\pscal{\Psi_n, D^V\Psi_n}\to \ell$
\item $\Psi_n\wto0$ weakly in $L^2(\R^3)$, that is, $\pscal{\Phi, \Psi_n}\to 0$ for all $\Phi\in L^2(\R^3)$.
\end{enumerate}
Then we can use this sequence to construct a spurious mode, by starting from any nice approximation basis. The idea is simply to add the vector $\Psi_{n}$ with $n\gg1$, to a given space $W_k$. The matrix of $D^V$ in the space ${\rm span}(W_k\cup\{\Psi_{n}\})$ becomes block-diagonal in the limit $n\to\ii$,
$$\begin{pmatrix}
(D^V)_{|W_k}& \simeq 0\\
\simeq 0 & \pscal{\Psi_{n}D^V\Psi_{n}}\simeq \ell
\end{pmatrix}.$$
The off-diagonal terms tend to zero due to the fact that $\Psi_n$ becomes asymptotically orthogonal to $D^V\Phi$, for any fixed $\Phi\in W_k$. One can therefore choose $n=n_k\gg1$ to have an eigenvalue as close to $\ell$ as we desire. In the limit $k\to\ii$, $\ell$ will be a spurious eigenvalue. 

So, we see that everything reduces to constructing sequences $\Psi_n$ satisfying the previous two conditions. This technique (and an improvement of it that is discussed later) was used in~\cite{LewSer-10} to study spurious modes for the Dirac operator. The results obtained in~\cite{LewSer-10} are summarized in the next section.

%%%%%%%%%%%%%%%%%
\section{Strategies to avoid Spurious Modes in Dirac Calculations}

The problem of spurious modes for the Dirac equation has a long history, starting with the celebrated computation of Drake and Goldman~\cite{DraGol-81} in a Slater-type basis set. Several solutions to avoid this phenomenon have been proposed in the literature~\cite{DraGol-81,Grant-82,Kutzelnigg-84,StaHav-84,DyaFae-90,Pestka-03,Shaetal-04,BouBou-10}. Our purpose here is to present the rigorous results which we have obtained in~\cite{LewSer-10} concerning the mathematical validity of these techniques. 

In the whole section we assume that $V$ is a potential that tends to 0 at infinity, and we systematically distinguish the case of $V$ being \emph{bounded} over the whole space $\R^3$, from \emph{attractive Coulomb-type} potentials. The latter means for us that there are finitely many points $R_1,...,R_M$ (the locations of the nuclei) at which $V$ behaves asymptotically like
$$V(\br)\underset{R\to R_m}{\sim} -\frac{\alpha Z_m}{|\br-R_m|},\quad \text{with}\ 0\leq \alpha Z_m\leq \frac{\sqrt{3}}2,$$
and that $V$ is bounded outside of these points $R_m$ (and tends to 0 at infinity). More general potentials can be considered, but we stick to the previous example for simplicity. We usually do not assume $V(\br)$ to have a specific sign.

There are two simple motivations for considering general potentials $V(\br)$ instead of just $V(\br)=-\alpha Z/r$. First, the potential of a finite-radius nucleus 
$$V(\br)=-\alpha Z\int_{\R^3}\frac{n(\br')}{|\br-\br'|}\,\dddr'$$
is always bounded if $n$ is a smooth function. Secondly, in practice $V(\br)$ is a self-consistent function containing both the (negative) nuclear and (positive) electronic potentials, the latter being smoother than the one of pointwise nuclei.

%%%%%%%%%%%%%%%%%%%%%%%%%%%%
\subsection{Pollution in upper/lower spinor basis}\label{sec:upper/lower}
%%%%%%%%%%%%%%%%%%%%%%%%%%%%

It is natural to use a basis which is made of upper and lower spinors, that is of functions of the form
$$\begin{pmatrix}\phi\\ 0\end{pmatrix}\text{ and } \begin{pmatrix}0\\ \chi\end{pmatrix}.$$
In the radial case~\eqref{eq:radial_basis}, this amounts to choosing two independent basis sets for the functions $u$ and $v$. It may be checked that a basis of this form never pollutes for the free Dirac operator $D^0$ and therefore one might think that it would not pollute for $D^0+V(\br)$. But this is actually not true, it is possible to get spurious modes even with a very nice bounded potential $V$.

\begin{theorem}[Pollution in upper/lower spinor basis~{\cite[Thm 2.7]{LewSer-10}}]\label{thm:upper-lower}
There exists an increasing sequence of spaces $W_n$ spanned by functions of the form
\begin{equation}
\begin{pmatrix}\phi\\ 0\end{pmatrix}\text{ and } \begin{pmatrix}0\\ \chi\end{pmatrix}, 
\label{eq:form_upper_lower}
\end{equation}
for which the intervals
\begin{equation}
\big[\max(-1,1+\inf(V))\,,\,1\big]\quad\text{and}\quad \big[-1\,,\,\min(1,\sup(V)-1]\big]
\end{equation}
are completely filled with spurious modes. This basis can be chosen to consist of gaussian functions multiplied by polynomials. 

There cannot be any spurious modes \emph{outside} of the above two intervals for a basis of the form~\eqref{eq:form_upper_lower}.
\end{theorem}

\begin{figure}[h]
\input{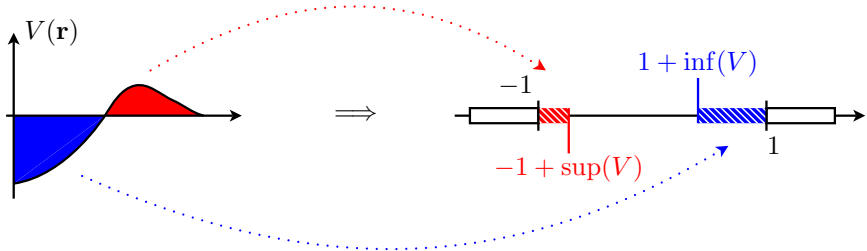}
\caption{Possible location of spurious modes in upper/lower spinor basis, depending on the size of the negative and positive parts of the external potential $V(\br)$ (Theorem~\ref{thm:upper-lower}).\label{fig:poll_upper_lower}}
\end{figure}

Note that since $V\to0$ at infinity by assumption, then we always have $\inf(V)\leq0$ and $\sup(V)\geq0$. For a negative potential $V$, the previous result says that we can fill the whole interval $[\max(-1,1+\inf(V)),1]$ with spurious modes. In the Coulomb case we have $\inf(V)=-\ii$, and therefore we can get spectral pollution everywhere in the gap. For a bounded potential $V$ such that $|V(\br)|\leq 2$, we can only get pollution in $[-1,-1+\sup(V)]\cup [1+\inf(V),1]$ (see Figure~\ref{fig:poll_upper_lower}). The result also says that spurious modes cannot appear outside of these intervals, but the minimax characterization of eigenvalues for Dirac operators proved in~\cite{DolEstSer-00} implies that the true eigenvalues indeed exactly lie in these intervals where pollution can occur. 

We conclude that choosing a basis made of upper/lower spinors can sometimes lead to spurious modes, if no further constraint is imposed. This is certainly well-known in the chemistry literature~\cite{DraGol-81}.

The proof of Theorem~\ref{thm:upper-lower} is intuitively easy. If we take an upper spinor, we get
$$\pscal{\begin{pmatrix}\phi\\0\end{pmatrix},(D^0+V)\begin{pmatrix}\phi\\0\end{pmatrix}}=\int_{\R^3}\big(1+V(\br)\big)|\phi(\br)|^2\dddr.$$
Recall that in our units $m=c=1$. Now we can make this converge to $1+V(\br_0)$, for any fixed $\br_0\in\R^3$ by choosing a sequence $\phi_n$ which gets more and more concentrated at this point, like a delta function. Such a sequence $\phi_n$ converges weakly to $0$ in $L^2(\R^3)$, hence we conclude from the discussion in Section~\ref{sec:method} that $1+V(\br_0)$ can be made a spurious eigenvalue for any $\br_0$ such that $V(\br_0)<0$. The same argument applied to lower spinors gives the result for the lower part of the gap.

%%%%%%%%%%%%%%%%%%%%%%%%%%%%
\subsection{Kinetic balance}\label{sec:KB}
%%%%%%%%%%%%%%%%%%%%%%%%%%%%

The most celebrated method used in practice to avoid spurious eigenvalues is the so-called \emph{kinetic balance}~\cite[Chap. 5]{Schwerdtfeger1}. It is implemented in all the quantum chemistry computer programs. The starting point is to write the eigenvalue equation as
$$
\left\{\begin{array}{l}
(mc^2+V)\phi+c\sigma\cdot(-i\nabla)\chi=(mc^2+\mu)\phi,\\[0.2cm]
c\sigma\cdot(-i\nabla)\phi+(-mc^2+V)\chi=(mc^2+\mu)\chi,
\end{array}\right.
$$
where we have re-introduced the speed of light $c$ for clarity. Here $\Psi=\begin{pmatrix}\phi\\\chi\end{pmatrix}$ is again written in the upper/lower component decomposition. Solving the second equation for $\chi$ gives
\begin{equation}
\chi=\frac{c}{2mc^2+\mu-V}\sigma\cdot(-i\nabla)\phi.
\label{eq:relation-chi-phi} 
\end{equation}
Of course this is not of great help since the eigenvalue $\mu$ is unknown \emph{a priori}. For $c\gg1$, however, we can hope that
$$\chi\simeq \frac{1}{2mc}\sigma\cdot(-i\nabla)\phi,$$
and this suggests to impose this relation between the basis for the upper spinor and that of the lower spinor. So, the kinetic balance method consists in choosing a basis $\phi_1,...,\phi_n$ for the upper spinor and taking the basis $\sigma\cdot\nabla \phi_1,...,\sigma\cdot\nabla \phi_n$ for the lower spinor~\cite{DraGol-81,Grant-82,Kutzelnigg-84,StaHav-84}.\footnote{Sometimes the basis is rather taken to be $\sigma_k\partial_k\phi_n$, which multiplies the number of lower spinors by 3.}

It is a common belief that the kinetic balance method is a useful tool to avoid spurious modes. The following theorem confirms this intuition for bounded potentials, but shows that the problem persists for Coulomb potentials.

\begin{theorem}[Pollution with kinetic balance~{\cite[Thm 3.4]{LewSer-10}}]\label{thm:KB}
If $V(\br)\leq 2$ is bounded from below, there is never any spurious mode in a kinetically balanced basis in 
$\big[\max(-1,\inf(V)+1)\,,\,1\big]$,
but there may be some in $\big[-1\,,\,\min(1,\sup(V)-1]\big]$. 

If $V$ is of Coulomb type, then there exists an increasing sequence of spaces $W_n$ spanned by functions of the form
$$\begin{pmatrix}\phi\\ 0\end{pmatrix}\text{ and } \begin{pmatrix}0\\ \sigma\cdot\nabla \phi\end{pmatrix},$$
for which there is pollution in the whole interval $[-1,1]$.
The basis can be chosen to consist of gaussian functions multiplied by polynomials.
\end{theorem}

\begin{figure}[h]
\input{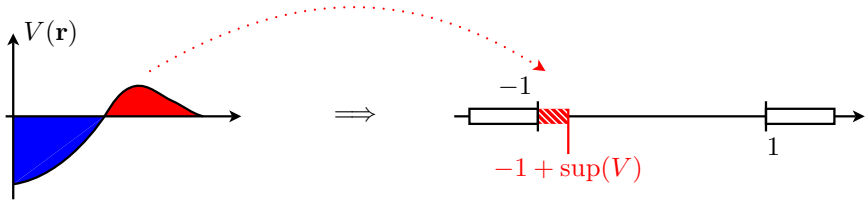}
\caption{Possible location of spurious modes in a kinetically balanced basis, for a \emph{bounded} potential $V(\br)$ (Theorem~\ref{thm:KB}). As compared to Figure~\ref{fig:poll_upper_lower}, the spurious modes corresponding to the attractive part of $V(\br)$ are suppressed. In a Coulomb potential, spurious modes can in principle fill completely the interval $[-1,1]$. \label{fig:poll_KB}}
\end{figure}

The theorem says that, in the case of bounded potentials, spurious eigenvalues are avoided in the upper part of the spectrum, but \emph{a priori} not in the lower part (Figure~\ref{fig:poll_KB}). This is because the kinetic balance condition is based on a non-relativistic limit for electrons in which the upper spinor is dominant. In particular, the result says that for negative bounded potentials, there will be no pollution at all.

On the other hand, the theorem says that, for Coulomb potentials, kinetic balance does not avoid the occurrence of spurious modes in general. Of course, this does not mean that they will necessarily show up in a given basis set, it only means that this is in principle possible.

We do not discuss here the proof that kinetic balance does not pollute for bounded potentials. The mathematical analysis is involved, and the interested reader should look at the details in~\cite{LewSer-10}. Rather, we quickly explain the strategy used in~\cite{LewSer-10} to prove the existence of spurious modes in the Coulomb case. The idea is very similar to that explained in Section~\ref{sec:method}. The main difference is that we cannot add only one vector to a given basis set, because we have to include both $(\phi_n,0)$ and its kinetically balanced counter part $(0,\sigma\cdot\nabla \phi_n)$. However, it is clear that if we can find a sequence $\phi_n$ such that
\begin{enumerate}
\item the $2\times2$ matrix of $D^0+V$ in the basis $\begin{pmatrix}\phi_n\\ 0\end{pmatrix}\,,\,\begin{pmatrix}0\\ \sigma\cdot\nabla\phi_n\end{pmatrix}$ has $\ell$ in its spectrum in the limit $n\to\ii$;
\item $\phi_n\wto0$ and $\sigma\cdot\nabla\phi_n\wto 0$ in $L^2(\R^3)$,
\end{enumerate}
then the argument is the same as in Section~\ref{sec:method}: The matrix of $D^V$ in $\{W_k$, $(\phi_{n_k},0)$, $(0,\sigma\cdot\nabla \phi_{n_k})\}$ is almost diagonal by blocks
$$\begin{pmatrix}
(D^V)_{|W_k}& \simeq 0\\
\simeq 0 & (D^V)_{|{{\phi_{n_k}}\choose 0},{0\choose {\sigma\cdot\nabla\phi_{n_k}}}}
\end{pmatrix}.$$
For $V=-\alpha Z/r$, the idea of~\cite{LewSer-10} is to take a contraction (that is, a linear combination) of two gaussians concentrated at the origin, where the Coulomb potential blows up:\footnote{Actually, in~\cite{LewSer-10}, the function is taken of the form $\phi_n=\left(f(nr^2)+g(\delta n r^2)\right){1\choose 0}$ where $f$ and $g$ are chosen with disjoint support, which simplifies some calculations.}
\begin{equation}
\phi_n=\left(e^{-nr^2}+\delta^{1/4}e^{-n\delta r^2}\right)\begin{pmatrix}1\\0\end{pmatrix}.
\label{eq:form_Lewin_Sere} 
\end{equation}
It is a tedious but simple calculation to verify that the $2\times2$ matrix of $D^V$ in the associated basis can have one eigenvalue lying in the gap $(-1,1)$, for any $n$ large enough, provided that $\delta$ is tuned appropriately. 

\begin{figure}[h]
\includegraphics[width=10cm]{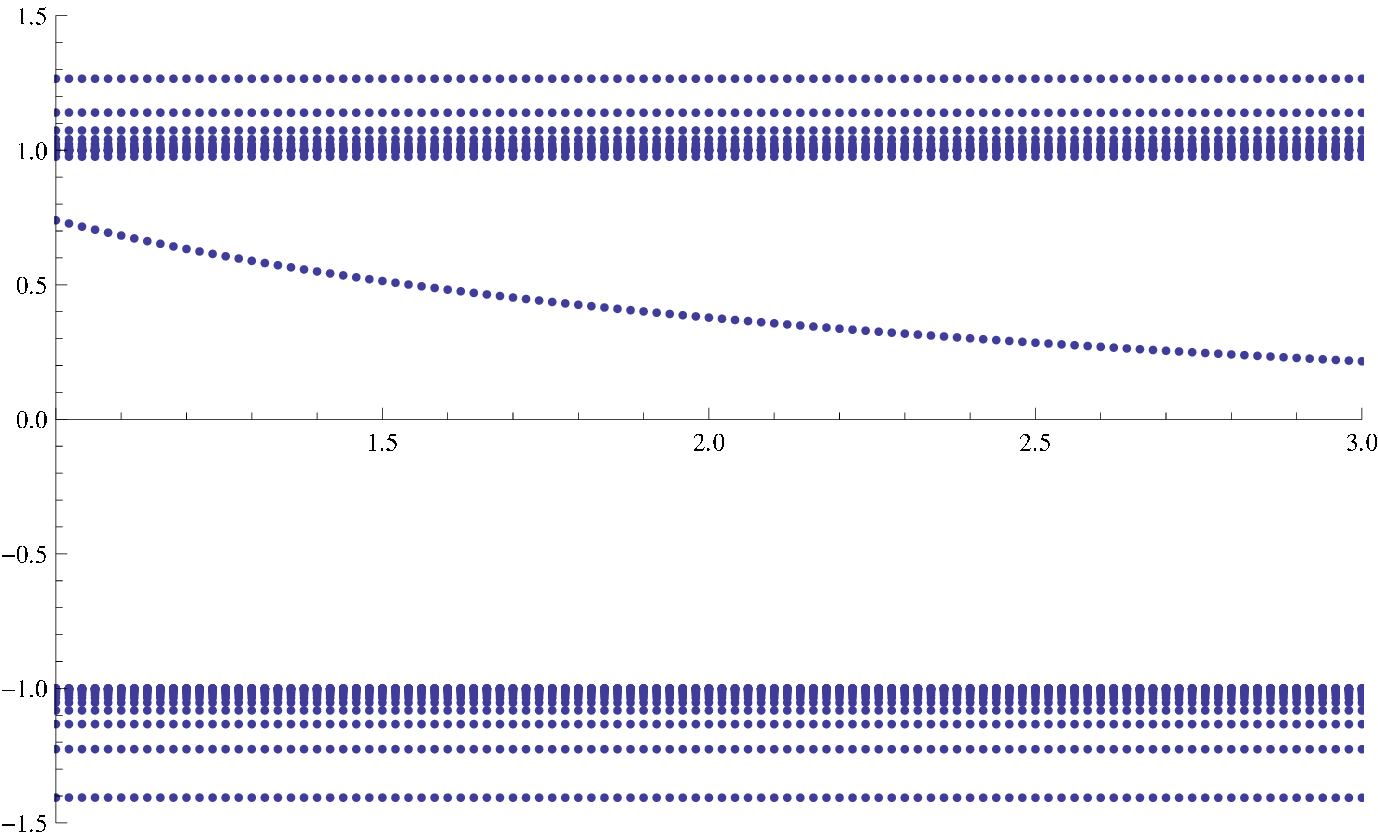}
\caption{Spectrum of $D^0-30\alpha/r$ computed in the kinetically-balanced basis set~\eqref{eq:basis_KB}, in terms of the parameter $10^{-4}\delta$.\label{fig:kinetic}}
\end{figure}

In Figure~\ref{fig:kinetic} we display the spectrum of $D^V$ in a (radial) kinetically balanced basis using for the upper component
\begin{equation}
e^{-a_1 r^2}\begin{pmatrix}1\\0\end{pmatrix}\,,\,...\,,\,e^{-a_n r^2}\begin{pmatrix}1\\0\end{pmatrix}\,,\, \left(e^{-b r^2}+\delta^{1/4}e^{-b\delta r^2}\right)\begin{pmatrix}1\\0\end{pmatrix}
\label{eq:basis_KB}
\end{equation}
where the $a_i$ are as before the gaussian parameters of the 6-31G basis set for zinc, $Z=30$, $b=10^6\alpha^2$ and where $\delta$ is varied in a neighborhood of $\sim 10^4$. Again we observe a clear spurious mode due to the additional test function~\eqref{eq:form_Lewin_Sere}.

%%%%%%%%%%%%%%%%%%%%%%%%%%%%
\subsection{Atomic balance}
%%%%%%%%%%%%%%%%%%%%%%%%%%%%

It is clear from the previous section that the occurrence of spurious modes in kinetically balanced basis sets is purely due to the singularity at zero of the Coulomb potential. This fact is also well-known to chemists~\cite{DyaFae-90,Pestka-03}. Taking into account this singularity amounts to modifying the kinetic balance condition at $0$. Indeed, for $r\ll1$, then~\eqref{eq:relation-chi-phi} rather becomes 
$$\chi(\br)\simeq \frac{c}{2mc^2-V(\br)}\sigma\cdot(-i\nabla)\phi(\br)$$
since $V(\br)$ can be very negative. This suggests to impose the relation (in units such that $m=c=1$)
$\chi_n=(2-V)^{-1}\sigma\cdot\nabla \phi_n$ for the lower spinor basis, a technique which is called \emph{atomic balance}. 

\begin{theorem}[Pollution for atomic balance~{\cite[Thm 3.5]{LewSer-10}}]
For $V\leq0$ a purely attractive bounded or Coulomb type potential, a basis constructed by the atomic balance method does not yield any spurious mode in the gap $(-1,1)$. If $V$ has a positive component, then one can still get spurious modes in the interval $\big[-1\,,\,\min(1,\sup(V)-1]\big]$. 
\end{theorem}

We see that the atomic balance condition allows to avoid spurious modes, even in the Coulomb case. This is of course at the cost of a higher numerical complexity, since the factor $(2-V)^{-1}$ will certainly raise some complications. The atomic basis method does not seem to have spread out much in quantum chemistry packages.

%%%%%%%%%%%%%%%%%%%%%%%%%%%%
\subsection{Dual kinetic balance}
%%%%%%%%%%%%%%%%%%%%%%%%%%%%

In the previous sections we have considered two possible methods (the kinetic and atomic balance) and we have explained in which situation these avoid spurious eigenvalues \emph{in the upper part of the spectrum}. These methods are based on a special relation between the upper and lower spinors in the non-relativistic limit, and they can only properly deal with electrons. They cannot help to avoid positronic spurious modes.

In this and in the following section, we consider two methods which are completely symmetric with respect to exchanges of electrons into positrons. The first is the so-called \emph{dual kinetic balance} method which was introduced by Shabaev et al in~\cite{Shaetal-04}. It consists in taking basis elements of the special form
\begin{equation}
\begin{pmatrix}
\phi\\ -i\varepsilon \sigma\cdot \nabla\phi\end{pmatrix}\quad\text{and}\quad\begin{pmatrix}
-i\varepsilon \sigma\cdot \nabla\phi\\ -\phi\end{pmatrix},
\end{equation}
see~\cite[Eq. (24)--(25)]{Shaetal-04}. In the original article, the parameter is $\varepsilon=1/(2mc^2)=1/2$ but we will keep it free here to emphasize its role. 

\begin{theorem}[Pollution with dual kinetic balance~{\cite[Thm 3.9]{LewSer-10}}]\label{thm:dual}
Let $0<\epsilon\leq1$. We can find an increasing sequence of spaces $W_n$ spanned by functions of the form
\begin{equation}
\begin{pmatrix}
\phi_n\\ -i\varepsilon \sigma\cdot \nabla\phi_n\end{pmatrix}\quad\text{and}\quad\begin{pmatrix}
-i\varepsilon \sigma\cdot \nabla\phi_n\\ -\phi_n\end{pmatrix},
\label{eq:form_DKB} 
\end{equation}
for which the intervals
$$
\left[\max\left(-1,1+2\left(\frac
{1}\varepsilon-1\right)+\inf(V)\right)\,,\,1\right]$$
and 
$$\left[-1\,,\,\min\left(1,\sup(V)-1-2\left(\frac
{1}\varepsilon-1\right)\right)\right]
$$
are completely filled with spurious modes.
The basis can be chosen to consist of gaussian functions multiplied by polynomials. However, there are no spurious modes outside of these two intervals in a basis of the form~\eqref{eq:form_DKB}. In particular, we can fill the gap $(-1,1)$ with spurious modes for Coulomb potentials.
\end{theorem}

\begin{figure}[h]
\input{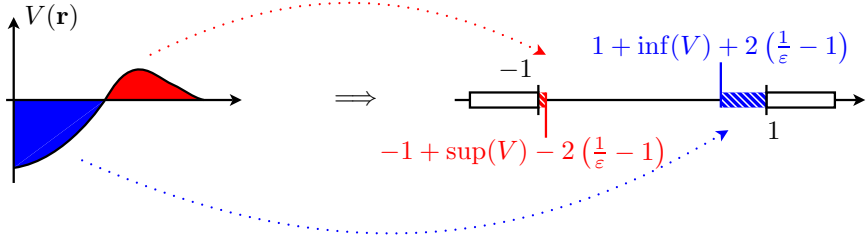}
\caption{Possible location of spurious modes in a dual kinetically balanced basis, for a bounded potential $V(\br)$ (Theorem~\ref{thm:dual}). As compared to Figure~\ref{fig:poll_upper_lower}, the two intervals where spurious modes can appear are shifted by the same amount $2(1/\epsilon-1)$.  \label{fig:poll_DKB}}
\end{figure}

We see that the dual kinetic balance behaves well in both the upper and lower parts of the gap, for bounded potentials, in the sense that the two intervals in which spurious modes can appear, are shifted by the same amount $2(1/\epsilon-1)$ (Figure~\ref{fig:poll_DKB}. In particular,  spurious modes will be completely avoided if
$$\epsilon\leq \frac{1}{2+|V(\br)|},$$
for all $\br$. Note that this is impossible for Coulomb potentials which are unbounded.

%%%%%%%%%%%%%%%%%%%%%%%%%%%%
\subsection{Absence of pollution in free basis}
%%%%%%%%%%%%%%%%%%%%%%%%%%%%

So far, we seem to have encountered no perfect method. The kinetic balance technique works well in the upper part of the spectrum for bounded potentials, but it is inefficient in the lower part. The atomic balance behaves better for attractive Coulomb potentials but the problem is not at all solved for the spurious modes associated with the positive component of the potential $V(\br)$. Finally, the dual kinetic balance method can be tuned to work for a bounded potential whatever its sign, but it is not adapted to Coulomb singularities.

We would like to present in this last section a method that works in all situations, independently of the sign of $V(\br)$ and of its local singularities. Of course, there is a price to pay and the numerical cost might be increased a lot. Nevertheless, it seems to not have been tested yet in practice and we would like to advertise it.

The idea is to use a basis that is adapted to the free Dirac operator $D^0$. In momentum space, the latter may be diagonalized as follows
$$\begin{pmatrix}
1 & \sigma\cdot \bp\\
\sigma\cdot \bp & -1
\end{pmatrix}=U(\bp)^*\begin{pmatrix}
\sqrt{1+p^2}\,\1_2 & \\
0 & -\sqrt{1+p^2}\,\1_2
\end{pmatrix}U(\bp)$$
where $U(\bp)$ is the unitary matrix
$$U(\bp)=\sqrt{\frac{1+(1+p^2)^{-1/2}}{2}}\1_4+\sqrt{\frac{1-(1+p^2)^{-1/2}}{2}}\beta\, \alp \cdot \frac{\bp}{p}.$$
The electronic states form an infinite-dimensional space defined as
$$\gH^+=\left\{\Psi\in L^2(\R^3,\C^4)\ \Big|\ U(\bp)\widehat{\Psi}(\bp)\in \text{span}\left\{\begin{pmatrix}
1\\0\\0\\0\end{pmatrix}\,,\,\begin{pmatrix}0\\1\\0\\0\end{pmatrix}
\right\}\right\}.$$
% Similarly, the positronic states are those in
% $$\gH^-=\left\{\Psi\in L^2(\R^3,\C^4)\ \text{such that}\ U(\bp)\widehat{\Psi}(\bp)\in \text{span}\left\{\begin{pmatrix}
% 0\\0\\1\\0\end{pmatrix}\,,\,\begin{pmatrix}0\\0\\0\\1\end{pmatrix}
% \right\}\right\}$$
There is a similar definition for the positronic space $\gH^-$ and the full Hilbert space is the direct sum of the previous two,
$L^2(\R^3,\C^4)=\gH^+\oplus\gH^-$.
The result is the following.

\begin{theorem}[Absence of pollution in free basis~{\cite[Thm. 2.10]{LewSer-10}}]
Let $V$ be a bounded or (repulsive or attractive) Coulomb-type potential. Consider a sequence of discretization spaces $W_n$ admitting a basis of functions, belonging either to $\gH^+$ or to $\gH^-$. Then there are never any spurious modes.
\end{theorem}

So if we use a basis which is adapted to the free Dirac operator $D^0$ in the sense that it only contains electronic and positronic free states, there is never any spurious eigenvalues. This result is intuitive because it is clear that such a basis cannot pollute when $V\equiv0$, and so one might expect that it also does not pollute for $V\neq0$. One has to be careful with such arguments. Recall the upper/lower spinor basis discussed in Section~\ref{sec:upper/lower} which never has spurious modes when $V\equiv0$ but may have some when $V\neq0$.

The main question is how to implement this in practice. If we have a given basis set, we could project it onto the electronic and positronic subspaces $\gH^\pm$, but this can only be done approximately. It is an interesting question to investigate which precision is necessary to avoid spectral pollution in a given sub-interval of the gap. No explicit error bounds are known and they would be very useful for the development of an efficient strategy in this direction.

%%%%%%%%%%%%%%%%%%%%%%%%%%%%
\section{Conclusion and open problems}
%%%%%%%%%%%%%%%%%%%%%%%%%%%%

In this paper we have considered several methods which can be used to avoid spurious modes when computing eigenvalues of Dirac operators, typically in a Coulomb potential. Our findings are summarized in Table~\ref{tab:summary} below.

\begin{table}[h]
\begin{tabular}{|l|c|c|c|}
\hline
& bounded $V\leq0$ & bounded $V\geq0$ & $\leq0$ Coulomb\\
\hline
upper/lower & {\color{red}\XSolidBold}& {\color{red}\XSolidBold}& {\color{red}\XSolidBold}\\
kinetic balance &  {\color{green}\CheckmarkBold}& {\color{red}\XSolidBold} & {\color{red}\XSolidBold}\\
atomic balance & {\color{green}\CheckmarkBold}& {\color{red}\XSolidBold} & {\color{green}\CheckmarkBold}\\
dual kinetic balance & {\color{green}\CheckmarkBold} & {\color{green}\CheckmarkBold} & {\color{red}\XSolidBold}\\
free basis & {\color{green}\CheckmarkBold} & {\color{green}\CheckmarkBold} & {\color{green}\CheckmarkBold}\\
\hline
\end{tabular}

\medskip

\caption{Summary of the results.\label{tab:summary}}
\end{table}

Let us emphasize that we have considered here the most pessimistic point of view. We are not able to say if spurious modes will appear in a given basis. We are only able to prove that spurious modes will \emph{never appear} for a certain class of methods, in a region of the spectrum or, on the contrary, to \emph{construct counterexamples} showing that pollution is possible with the given constraints. The counterexamples may of course seem to be \emph{ad hoc} but they already give a hint of the possible problems that may arise in practical calculations.

It is a widely open problem to find simple criteria which could be applied to a given basis set, instead of a whole class of basis sets as we did here. For gaussians, one may think of a criterion in phase space which would measure how the latter is progressively filled up. Our counterexamples are always based on spatially very spread-out or very concentrated functions, which would look completely isolated from the other elements of the basis in phase space. Turning this intuition into a rigorous statement is an interesting open problem.

\bigskip

\small\noindent \textbf{Acknowledgement.} M.L. would like to thank Lyonell Boulton and Nabile Boussaid for stimulating discussions, in particular concerning the numerical experiments of this article. M.L. has received financial support from the European Research Council under the European Community's Seventh Framework Programme (FP7/2007-2013 Grant Agreement MNIQS 258023). M.L. and \'E.S. acknowledge financial support from the French Ministry of Research (ANR-10-BLAN-0101).

% \bibliographystyle{siam}
% \bibliography{biblio}

\end{document}